\documentclass[journal=jacsat,manuscript=article]{achemso}

\usepackage[version=3]{mhchem} 
\author{Ping-Chun Chen}
\author{Mashnoon Alam Sakib}
\author{Mariia Stepanova}
\author{\newline Melika Momenzadeh}
\affiliation[University of California, Irvine]
{Department of Electrical Engineering and Computer Science, University of California, Irvine, California 92697, United States}
\author{Maxim R. Shcherbakov}
\affiliation[University of California, Irvine]
{Department of Materials Science and Engineering, University of California, Irvine, California 92697, United States}
\email{maxim.shcherbakov@uci.edu}

\title[An \textsf{achemso} demo]
  {Multifunctional Barophotonic Control of Resonators and Metasurfaces}

\abbreviations{IR,NMR,UV}

\begin{document}




\begin{abstract}

Actively tunable nanophotonic platforms that control light–matter interactions enable reconfigurable optical systems and programmable photonic integrated circuits. Hydrostatic pressure provides a noninvasive and material-agnostic mechanism for modulating the refractive index and resonance conditions without introducing free carriers or structural damage. Here, we demonstrate multiple pressure-dependent functionalities in silicon nitride nanostructures, including resonance tuning, refractive index modulation, and polarization state conversion. Applying a pressure of up to 5~GPa, we observe a Fabry-P\'erot resonance shift of up to 30~nm and a relative refractive index decrease of up to $4\%$. Based on the results, we design and examine, to the best of our knowledge, the first extreme-pressure-tunable, polarization-converting metasurface, which tunes the ellipticity and orientation angle of the output light. These findings establish pressure-controllable silicon nitride as a viable platform for reconfigurable photonics and extreme-environment nanophotonic systems, including deep-ocean exploration, planetary interiors, and space applications.


\end{abstract}

Photonic resonators (PRs)---passive devices such as ring resonators, photonic crystals, Fabry-Pérot (FP) cavities, and metasurfaces---drive light-matter interactions at select frequencies and play an increasingly crucial role in modern telecommunications\cite{W.Liang, PabloMarin-Palomo, Attila, QiangXiao, HanQingYang, SiRanWang}, sensing\cite{HanQingYang, JunMa, SiRanWang}, and quantum technologies \cite{LukaszDusanowski, I-TungChen, GregorBayer,ChristianLeefmans}. However, the functionalities of most PRs are fixed after fabrication, limiting the scope of their capabilities and applications. Dynamically controllable PRs utilize external stimuli, such as electrical \cite{DavidGeorge, GoSoma, AgostinoDiFrancescantonio, PingYu, MaximR.Shcherbakov}, mechanical \cite{ChenZhang, JuYoungKim}, chemical \cite{HailongLiu, JialongPeng}, and opto-thermal effects \cite{XingboLiu, JunghyunPark, ZhihuaZhu, M.Wuttig}, to tune the refractive index of PRs' constituents, resulting in 
phase, polarization, resonance wavelength, and spectral lineshape manipulation. Tunable PRs can 
reduce the number of devices in photonic integrated circuits, adapt to environmental changes, and enable real-time reconfiguration, 
generating growing interest in their implementations in programmable and reconfigurable optical systems \cite{WeilinLiu, WimBogaerts, XingyuanXu, LoubnanAbou-Hamdan, YasirSaifullah, ChenshengLi, ZhengXingWang}. 

Hydrostatic pressure remains one of the underexplored routes to tunable PRs. Extreme, gigapascal-scale hydrostatic pressure can promote changes in electronic bandgaps \cite{Y.Gao, S.Huang, LinglongZhang, AzkarSaeedAhmad}, light emission \cite{L.Zhao, YadongHuang, JiaxiangWang, ChunyanJiang}, and phase transitions \cite{S.Huang, A.Kundu, LinglongZhang, XuqiangLiu, SizhanLiu}, leading to pronounced changes in optical properties 
at room temperature. The pressure-tunable refractive index has been utilized to tune nanolaser modes\cite{YadongHuang}, FP resonances\cite{MgO, diamond-SiC-cBN, Ge, ZnO, GaP}, and surface plasmon resonances\cite{CaminoMartin-Sanchez}, leading to applications in sensing \cite{MarcinRunowski, JiapengZheng}.
However, simultaneous pressure-induced control over multiple functionalities of the same device has yet to be demonstrated. 

Here, we propose  multifunctional silicon-nitride-based barophotonic resonators 
that simultaneously control the intensity and polarization state of light with 
hydrostatic pressure $P$. 
We apply up to $P=5$~GPa to metal-coated thin-film FP resonators and observe 
up to $\Delta \lambda_{\rm FP}=30$~nm in FP resonance blueshift 
and a 
SiN refractive index variation $\Delta n/n=0.04$
, verifying strong pressure tunability of amorphous silicon nitride. 
We then demonstrate 
a multifunctional 
metasurface-based 
device 
that leverages the pressure-sensitive refractive index and anisotropy to tune its resonances 
and the polarization state of transmitted light at the same time
. In this device, under $P=1.85$~GPa, we observe $\Delta \lambda_{\rm W}=5.5$~nm and $\Delta \lambda_{\rm SP}=14 $~nm shifts of the 
waveguide mode and surface plasmon resonance, respectively. 
Pressure-induced changes in orientation $\Delta\psi$ and ellipticity $\Delta\chi$ of the polarization state of light reach $7.3^\circ$ and $7.4^\circ$, respectively, demonstrating 
a pressure-tunable polarization state converter capability. 
The 
experimental data agree with finite-difference time-domain (FDTD) simulations, which reveal the dominating role of SiN in the tunability of FP resonances and a competing nature of contributions from the pressure medium and SiN in the metasurface-based PR. 
Our results establish barophotonic resonators as a new pathway toward multifunctional and reconfigurable photonic systems.

\begin{figure}[t]
  \centering
  \includegraphics[width=1\linewidth]{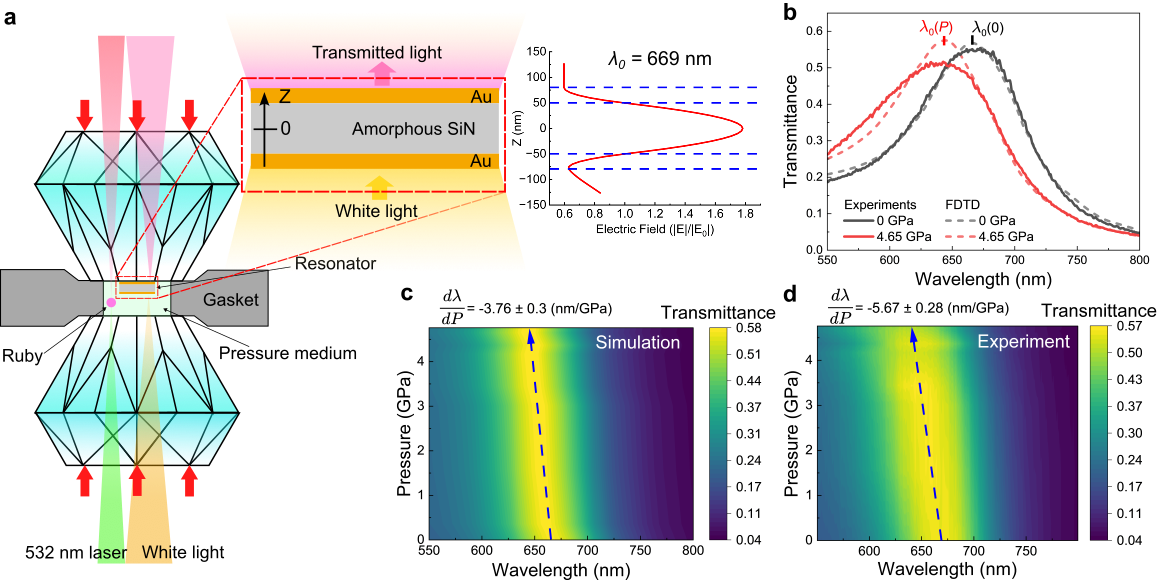}
  \caption{SiN Fabry–P\'erot resonators  under high pressure. \textbf{a,} Schematic side view of the high-pressure diamond anvil cell setup, where the hole in the gasket between the diamonds forms the high-pressure chamber. The Fabry–P\'erot resonator sample was illuminated by white light, while ruby was excited by a 532 nm laser and functions as a pressure indicator. By forcing the diamonds inward, pressure is applied to the sample through the pressure medium. The structure of the Fabry–P\'erot resonator is illustrated in the inset, showing a $l = 100$~nm thick SiN sample coated with $t = 30$~nm  gold on both sides and exposed to the white light source. The magnitude of the resonator electric field along the $z$-axis is plotted next to the structure (the blue dashed lines indicate the gold coating), demonstrating the lowest-order Fabry–P\'erot resonance at $\lambda = 669$ nm at ambient pressure. \textbf{b,} Comparison of simulation results with experimental data at ambient pressure and high pressure $P$ = 4.65~GPa. The resonance blueshifts up to 30 nm by increasing the applied pressure from ambient pressure to $P$ = 4.65~GPa. \textbf{c–d,} The FDTD simulation and experiment of SiN Fabry–P\'erot transmittance as a function of pressure, respectively.}
  \label{fig:Figure 1}
\end{figure}

Figure \ref{fig:Figure 1}a shows the diamond anvil cell (DAC) setup and the sample configuration. The sample within the DAC was illuminated by an incandescent light source for resonance measurements 
and a 532~nm continuous-wave (CW) solid-state laser serving as the excitation source for ruby photoluminescence (PL) measurements used for pressure calibration. The complete optical setup 
is shown in Supporting Information Figure S3.
We fabricated FP resonators using commercially available amorphous SiN membranes with a thickness of \textit{l} = 100 nm by depositing $t = 30$~nm of gold on both sides, as shown in the inset of Figure \ref{fig:Figure 1}a.
The inset also shows the FDTD-calculated magnitude of the electric field in the resonator excited by a normally incident plane wave at a wavelength of $\lambda = 669$~nm
, where the blue dashed lines indicate the gold coatings, confirming that a first-order FP mode is confined within SiN. For more details on sample preparation, DAC, and optical setup, refer to Supporting Information Sections 1–3.

The results of the FDTD simulation (dashed lines) and the experimental SiN FP transmittance (solid lines) at ambient pressure and $P=4.65$~GPa are shown in Figure \ref{fig:Figure 1}b in black and red, respectively. 
To precisely determine the peak positions of the resonance and their shifts under pressure, a second-order polynomial function was implemented to fit the experimental results. We observed a blueshift of the resonance by up to 30 nm when the applied pressure increased from ambient pressure to 4.65~GPa.
We note that the difference between spectral lineshapes in FDTD and experiment under 4.65~GPa might result from the wrinkles generated during sample transfer, which become enlarged under pressure (see Supporting Information Figure S5).

The FDTD and experimental results of transmitted spectrum at different pressures are illustrated in Figure \ref{fig:Figure 1}c and Figure \ref{fig:Figure 1}d, respectively.
The derivatives of peak position with respect to pressure are $\frac{d\lambda(P)}{dP}$ = $-3.76\, \pm\, 0.3~\mathrm{nm/GPa}$ and $\frac{d\lambda(P)}{dP}$ =  $-5.67\,\pm\,0.28~\mathrm{nm/GPa}$ for the simulation and experiment, respectively, which are significantly larger than that of ruby ($\frac{d\lambda(P)}{dP}$ = 0.36 nm/GPa). This high pressure sensitivity demonstrates SiN PR's potential for use as a pressure sensor.

\begin{figure}[ht]
  \centering
  \includegraphics[width=1\linewidth]{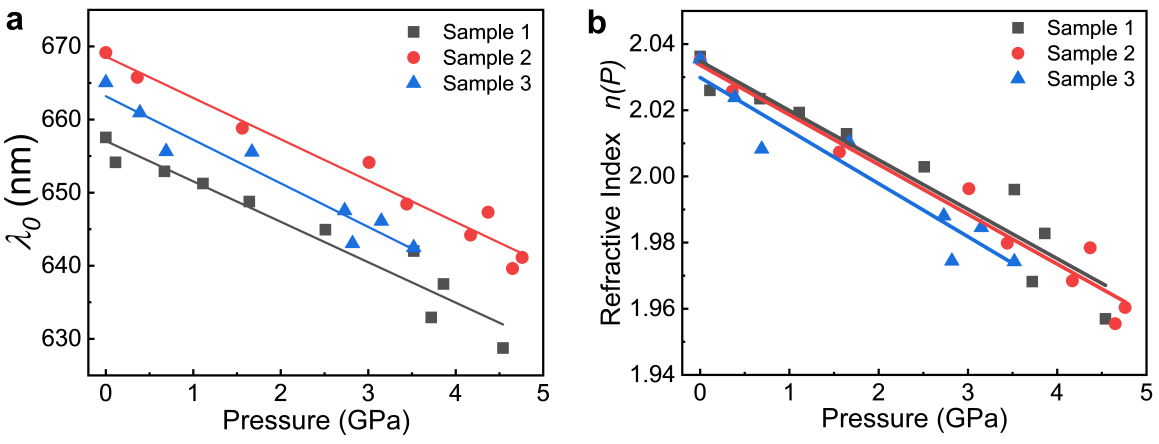}
  \caption{Refractive index tunability of SiN under pressure. \textbf{a,} Peak position of the first-order mode of the FP resonance shifts as a function of pressure. The squares, triangles, and circles represent different samples, and the solid lines are linear fits. For all samples, the peak positions exhibit blueshifts of up to 30~nm with increasing pressure up to 5~GPa. \textbf{b,} The refractive index of SiN, $n(P)$, as extracted using Eq.~(\ref{RI_equation}). 
  A decrease of 4\% in $n(P)$ under 5~GPa of applied pressure 
  is consistent across different samples.}
  \label{fig:Figure 2}
\end{figure}

Figure \ref{fig:Figure 2}a demonstrates the FP resonances 
as a function of pressure, exhibiting a consistent blueshift under increasing pressure.
We note that the slight difference in wavelength at ambient pressure (0~GPa in Figure 2a) among the samples might result from the fabrication tolerances in thickness and roughness of the gold coatings.

To understand the pressure dependence of SiN's refractive index, we examined the relationship between the refractive index, the resonance shift, and the thickness of the cavity \cite{diamond-SiC-cBN}: 
\begin{equation}
        \label{RI_equation}
        \frac{n(P)}{n(0)} = \frac{\lambda_m (P)}{\lambda_m (0)} \frac{l(0)}{l(P)},
\end{equation}
where $n$ is the refractive index, $\lambda_m$ is the wavelength of the fringe with order $m$ ($m =0$ in our case, see the inset in Figure \ref{fig:Figure 1}a), and $l$ is the thickness of the cavity.
The thickness variation under pressure was calculated using the Murnaghan equation of state:
$\frac{l(P)}{l(0)} = \left( 1 + \frac{B'_0 P}{B_0} \right)^{\scriptscriptstyle -\frac{1}{3B'_0}},$
where $B_0$ and $B'_0$ are the bulk modulus and its pressure derivative
, taken as $B_0=290$~GPa and $B_0'=4$ for SiN.\cite{Youngs_Modulus, Derivative_of_Modulus} 
The thickness variation of SiN, from ambient condition to nearly 5~GPa, is only about 0.55\% (see Supporting Information Section S7), whereas the resonance shift exceeds 4.5\%. 
Using Eq.~(\ref{RI_equation}), the refractive index of SiN as a function of pressure is plotted in Figure \ref{fig:Figure 2}b. The refractive index $n$ decreases linearly by around 4\% with increasing pressure up to 5~GPa.
This result indicates that the refractive index variation dominates the resonance shift over the thickness variation under pressure.
The contribution of gold coating to the pressure-induced resonance shift is negligible (see Supporting Information Section S7).
To the best of our knowledge, this is the first report of SiN under pressure, and the refractive index variation exhibits a pressure-dependent trend similar to that reported in other materials (see Supporting Information Table I).
This behavior might be attributed to the compression of SiN’s bond structure, which strengthens the covalent bonds and decreases the local polarizability of the material, thereby decreasing the refractive index, similar to previous studies in other materials \cite{ZnO, SnO2_DFT, diamond-SiC-cBN}.

We calculated the refractive index pressure coefficient of SiN to be $\frac{dn}{dP} = -1.5\times 10^{-2}$ $\text{GPa}^{-1}$. To compare with another common tuning mechanism, the typical thermo-optic coefficient of SiN is $\frac{dn}{dT}=2.45 \times 10^{-5}~^\circ\text{C}^{-1}$\cite{Amir.Arbabi}, and the thermal decomposition temperature of SiN is $1750^\circ\text{C}$\cite{H.D.Batha}. The maximum achievable variation in the refractive index of SiN from room temperature to decomposition temperature is 2\%. In contrast, the pressure-tuning approach enables a larger refractive index modulation in SiN, providing a promising potential for pressure-controllable SiN optical platform and applications.

\begin{figure}
  \centering
  \includegraphics[width=1\linewidth]{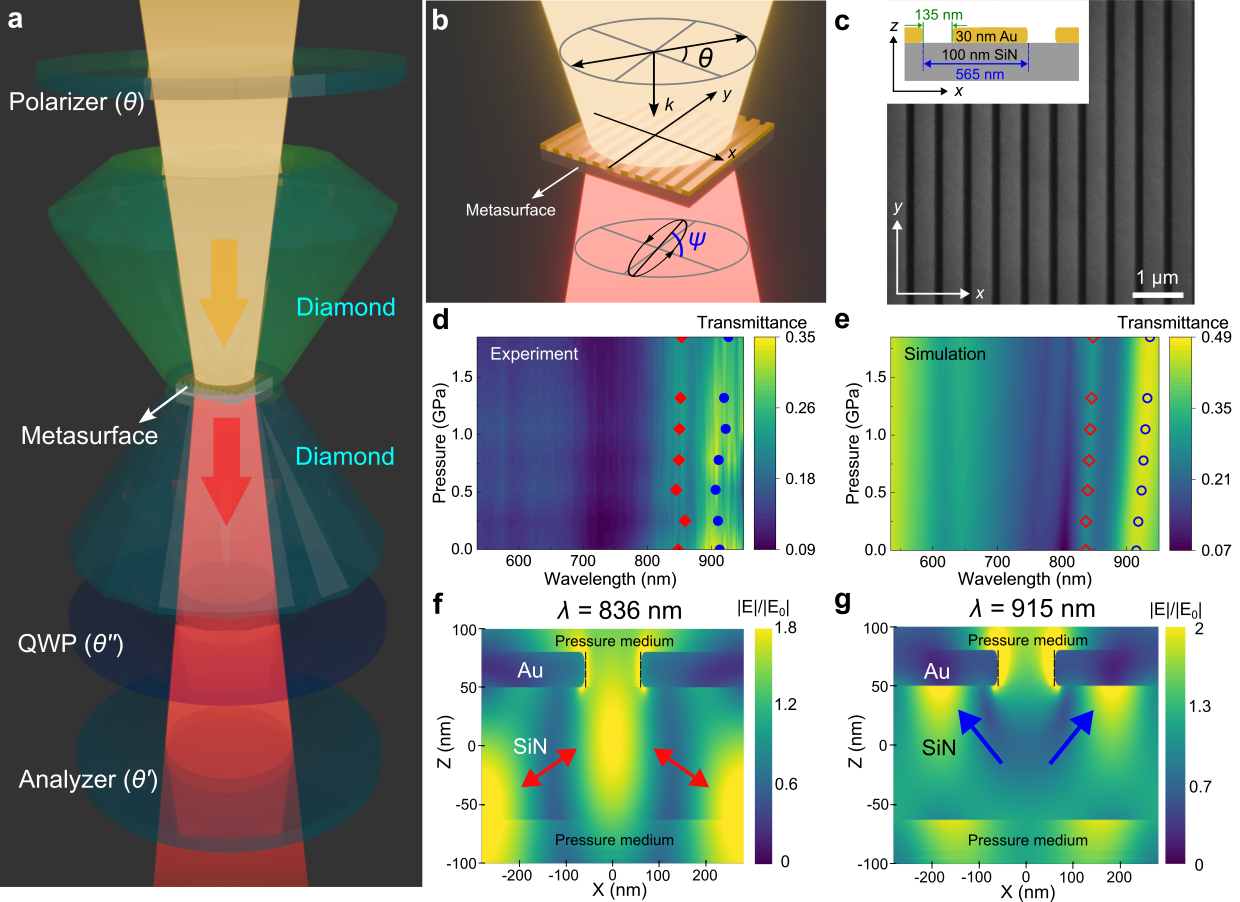}
  \caption{The configuration and the resonance tunability of a multifunctional barophotonic metasurface. \textbf{a,} The configuration of the metasurface placed in the DAC setup along with the optical components used for analyzing polarization states. The polarizer placed at $\theta = 45 ^\circ$ in front of the DAC sets the polarized state of incident light to be diagonal with respect to the $x-$axis of the sample frame. A quarter-wave plate (QWP) and analyzer are used to measure the Stokes parameters of the transmitted light. $\theta$, $\theta'$ and $\theta''$ are the angles of the polarizer, the analyzer, and the QWP, respectively, with respect to the $x-$axis of the sample frame. \textbf{b,} The schematic of the grating metasurface. Due to the anisotropy, the metasurface enables conversion of the incident light with linear polarization state at $\theta$ to an elliptical polarization state with orientation angle at $\psi$. \textbf{c,} Scanning electron microscope (SEM) image of the grating metasurface. 
  The inset shows the fabricated metasurface with a $t=30$~nm gold coating on top of $l=100$~nm SiN, a grating gap of 135 nm and a period of 565 nm. 
  \textbf{d–e,} The transmittance of metasurface under increasing pressure from experiment and simulation, respectively. The red diamonds and the blue circles represent the two resonant peaks, with solid markers for experiment and hollow ones for simulation. Both resonances exhibit redshifts of 5.5~nm and 14~nm at $P=1.85$~GPa in experiment, and 11~nm and 21~nm in simulation.
  \textbf{f} and \textbf{g} are the electric fields of the resonances at $\lambda = 836$~nm and $\lambda = 915$~nm, respectively, illustrating a waveguide mode (indicated by red arrow) and the surface plasmon resonance (indicated by blue arrow). 
  }
  \label{fig:Figure 3}
\end{figure}

We fabricated and characterized a grating metasurface (MS) 
that exhibits simultaneous tunability in its resonance and the polarization state of transmitted light, leveraging its pressure-dependent refractive index and anisotropy.
Figure \ref{fig:Figure 3}a shows the configuration of the MS placed in the DAC setup along with the optical components used for analyzing polarization states, and Figure \ref{fig:Figure 3}b illustrates the schematic of the grating MS. 
Due to the anisotropic structure, a linear polarization state of the incident light is converted into an elliptically polarized state after transmission through a grating MS. 
To analyze the polarization states of the transmitted light, we placed a linear polarizer at an angle $\theta = 45^\circ$ with respect to the $x$-axis of the sample frame after the incandescent light source to generate diagonally polarized incident light. An analyzer (a linear polarizer) at $\theta'$ and a quarter-wave plate (QWP) at $\theta''$ were placed in front of the detector to measure the Stokes parameters of the transmitted light. 
Details of the Stokes parameter measurements and the pressure estimation for the MS experiment are provided in the Supporting Information Section S3 and S8.

Figure \ref{fig:Figure 3}c shows the scanning electron microscope (SEM) image of the MS, and the inset illustrates the side view of the MS. 
The MS was prepared with a 30~nm gold coating on top of SiN, 
and the grating structure was etched into gold with a focused ion beam. 
The measured width of the grating gap is 135~nm, the period is 565~nm, and  the total area of the MS is 80 $\mu$m $\times$ 80 $\mu$m.

Figures \ref{fig:Figure 3}d and \ref{fig:Figure 3}e show the transmittance spectra of the MS under different pressures from experiment and simulation, respectively. 
The details of the MS simulation are provided in Supporting Information Section S4 and S10.
The red diamonds and the blue circles represent the two resonant peaks---a waveguide mode and a surface plasmon resonance---with solid markers for experiment and hollow ones for simulation.
We observe $\Delta \lambda_{\rm W}=5.5$~nm and $\Delta \lambda_{\rm SP}=14$~nm at $P = 1.85$~GPa in the experiment, and $\Delta \lambda_{\rm W}=11$~nm and $\Delta \lambda_{\rm SP}=21$~nm in simulation.
The derivatives of the peak positions of the waveguide mode and the surface plasmon resonance with respect to pressure, obtained from linear fits, are $\frac{d\lambda_{\rm W}(P)}{dP} = 1.08\, \pm\, 3.05~\mathrm{nm/GPa}$ and $\frac{d\lambda_{\rm SP}(P)}{dP} = 9.36\,\pm\,2.88~\mathrm{nm/GPa}$ in experiment. 
In simulation, the waveguide mode shows $\frac{d\lambda_{\rm W}(P)}{dP} = 6.62\, \pm\, 0.51~\mathrm{nm/GPa}$ and the surface plasmon resonance shows $\frac{d\lambda_{\rm SP}(P)}{dP} = 11.64\,\pm\,0.75~\mathrm{nm/GPa}$, which shows a trend similar to the experimental results. 
We note that the smaller experimental slope of $\lambda_{\rm W}(P)$ is skewed by the fitting result at $P=0.25$~GPa.

Figures \ref{fig:Figure 3}f and \ref{fig:Figure 3}g show the corresponding simulated electric fields of the resonant peaks at $\lambda = 836$~nm and $\lambda = 915$~nm, respectively.
At $\lambda = 836$~nm, a waveguide mode appears between the SiN and the pressure medium, as indicated by the red arrows. 
In contrast, a surface plasmon resonance is enhanced on the gold coating at $\lambda = 915$~nm, as indicated by the blue arrows.
As shown in Figure \ref{fig:Figure 3}f and \ref{fig:Figure 3}g, the pressure medium is involved in both resonances, demonstrating that the resonances are affected by both SiN and the pressure medium. We accounted for the pressure dependence of the pressure medium’s refractive index in our model as follows:\cite{CaminoMartin-Sanchez, JonH.Eggert}
\begin{equation}
        \label{Pressure Medium}
        n = n_0\left(\frac{P\alpha}{\beta}+1 \right)^{\scriptscriptstyle -\frac{1}{\alpha}},
\end{equation}
where  $n_0 = 1.3274$ is the pressure medium refractive index, and $\alpha = 21.3$ and $\beta = 14.3$~GPa are the empirical fitting parameters\cite{CaminoMartin-Sanchez, JonH.Eggert}.
By implementing both Eq.~(\ref{RI_equation}) for SiN and Eq.~(\ref{Pressure Medium}) for the pressure medium’s refractive indices into our calculation, we found that the pressure medium up to $P = 1.85$~GPa plays a greater role in the resonant behavior of the designed MS than SiN.



\begin{figure}
  \centering 
  \includegraphics[width=0.975\linewidth]{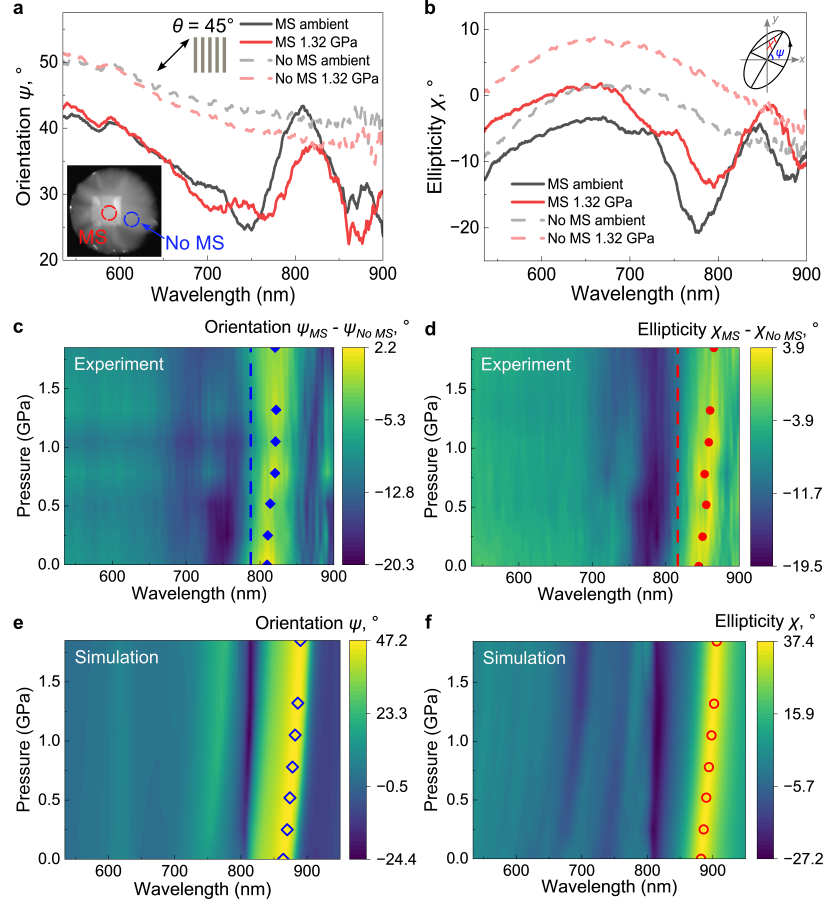}
  \caption{Polarization state tunability of the barophotonic metasurface. 
  \textbf{a-b,} The orientation $\psi$ and ellipticity $\chi$ of light transmitted through metasurface (`MS', solid lines) and without metasurface (`No MS', the dashed lines) at ambient pressure and at 1.32~GPa, respectively. The inset shows the measured `MS' and `No MS' regions in the DAC. \textbf{c-d,} Experimental differentials $\psi_{\text{MS}} - \psi_{\text{\textit{NoMS}}}$ and $\chi_{\text{\textit{MS}}} - \chi_{\text{\textit{NoMS}}}$ as functions of wavelength and pressure. 
  \textbf{e-f,} Simulated $\psi$ and $\chi$ as functions of wavelength and pressure. 
  Solid (experiment) and hollow (FDTD) symbols mark peak positions, showing 10 nm and 20 nm redshifts in \textbf{c} and \textbf{d}, and 25 nm in \textbf{e} and \textbf{f}. Dashed lines indicate wavelengths of maximum pressure-induced changes in $\psi$ and $\chi$: $\lambda =$ 787 nm with $\Delta\psi = 7.3^\circ$ and 816 nm with $\Delta\chi = 7.4^\circ$ in experiment.}
  \label{fig:Figure 4}
\end{figure}

The polarization-tuning functionality of the MS under pressure is shown by imaging the output of a Ø25 $\mu\text{m}$ multimode fiber on an area with the MS (the red dashed circle) and without the MS (No MS, the blue dashed circle), as shown in the inset optical image in Figure \ref{fig:Figure 4}a.
The orientation $\psi$ and ellipticity $\chi$ of the polarization state of the transmitted light through `MS' and `No MS' regions are calculated using the reconstructed Stokes parameters (see Supporting Information Section S3) and are plotted in Figure \ref{fig:Figure 4}a and \ref{fig:Figure 4}b, respectively, at ambient pressure and at $P = 1.32$~GPa. 
The smooth orientation $\psi$ and ellipticity $\chi$ in the `No MS' region across the spectra indicate no significant change to the polarization state of incident light with $\theta = 45^\circ$ after passing through the `No MS' area. 
On the contrary, $\psi$ and $\chi$ change significantly with MS, where the peak positions of $\psi$ and $\chi$ in the MS region redshift by $\sim 10$ and $\sim 15$ nm, respectively. 
We note that $\Delta\psi\sim 10^\circ$ and $\Delta\chi\sim 10^\circ$ in the No MS region across all wavelengths under pressure, indicating that the polarization state of the transmitted light might be affected by the environment. Therefore, to isolate the influence of the MS from the environment, we subtract both $\psi$ and $\chi$ of the `No MS' from those of the `MS' region and plot them in Figure \ref{fig:Figure 4}c and \ref{fig:Figure 4}d as a function of wavelength and pressure, respectively. 

The peak positions of the orientation $\psi$ and ellipticity $\chi$, represented by the blue diamonds and red circles in Figure \ref{fig:Figure 4}c and \ref{fig:Figure 4}d, respectively, show red shifts by $\Delta\lambda_\psi = 10$~nm and $\Delta\lambda_\chi = 20$~nm.
The blue dashed line in Figure \ref{fig:Figure 4}c 
indicates the maximum pressure-induced orientation change, $\Delta\psi \equiv \psi(\lambda, P_{\text{max}}) - \psi(\lambda, P_{\text{min}}) = 7.3^\circ$ at $\lambda = 787$~nm.
Figure \ref{fig:Figure 4}d shows the ellipticity $\chi$ varying from $3.9^\circ$ to $-19.5^\circ$, demonstrating the polarization handedness converting from right to left.
Similarly, the red dashed line 
shows the maximum pressure-induced ellipticity variation, $\Delta\chi\equiv\chi(\lambda, P_{max})-\chi(\lambda, P_{min}) = 7.4^\circ$ at $\lambda = 816$~nm.
These findings demonstrate that this pressure-tunable MS is able to work as a polarization converter. Furthermore, due to the wavelength-dependent sensitivity of pressure-induced polarization changes, the MS can act as a switch for the output polarization by wavelength selection.

Figure \ref{fig:Figure 4}e and \ref{fig:Figure 4}f show the orientation $\psi$ and ellipticity $\chi$ from FDTD, respectively. We note that the difference in peak positions in $\psi$ and $\chi$ between FDTD and experiment may be caused by fabrication tolerances and the wrinkles on the MS during sample transfer onto the diamond.
Both $\psi$ and $\chi$ redshift by $\sim 20$ nm under pressure, as depicted by the blue hollow diamonds and red hollow circles in Figure \ref{fig:Figure 4}e and \ref{fig:Figure 4}f, which aligns with the experimental results. The orientation $\psi$ rotates from $47.2^\circ$ to $-24.4^\circ$, and the ellipticity $\chi$ varies from $37.4^\circ$ to $-27.2^\circ$, demonstrating the polarization state of light converting from nearly right-handed circular polarization to left-handed polarization. 
These results are qualitatively consistent with the experimental observations, and the pressure-induced tunability demonstrates that the SiN MS can function as a pressure-controllable wave plate, offering a promising platform in dynamic optical systems.

The demonstration of a pressure-actuated polarization-controlling SiN metasurface highlights a new route toward tunable metasurfaces. The current grating metasurface can be further optimized to expand the modulation range of polarization states under pressure tuning to cover all polarization states, and to improve the quality factor of the resonance. Beyond polarization and resonance control, this pressure-actuated platform may enable follow-up studies of a broader range of metasurface functionalities, including phase, amplitude, and wavefront control.
Future investigation using density functional theory calculations and synchrotron X-ray absorption spectroscopy will be helpful to fully understand the variation of SiN's electronic structure and refractive index under pressure. 

In summary, we have demonstrated barophotonic control over resonant SiN photonic nanostructures and metasurfaces with multiple functionalities in an extreme pressure system. The resonance of the Fabry–P\'erot resonator blueshifts by around 30~nm within 5~GPa, and the resonance pressure sensitivity is $\frac{d\lambda}{dP} = -5.67 \pm 0.28$~nm/GPa, which is  larger than that of ruby, the standard pressure calibration material, by one order of magnitude. The total variation of SiN’s refractive index is shown up to $\Delta n/n \approx 4\%$. 
In an application example, we design and examine a grating metasurface structure acting as a pressure-controllable polarization state converter. 
The maximum pressure-induced orientation rotation and ellipticity variation reach, $\Delta\psi=7.3^\circ$ and $\Delta\chi=7.4^\circ$, respectively. These results demonstrate the pressure-tunability of SiN in nanoresonators and metasurfaces in resonance tuning, refractive index variation, and polarization state conversion, showing the potential of pressure-controllable SiN photonic devices for advanced and programmable photonic systems. Moreover, the demonstration of pressure-controlled nanophotonics highlights the feasibility of nanophotonic operation under extreme conditions, which may be relevant for photonic systems in high-pressure or harsh environments, such as those encountered in deep-ocean, planetary, or space settings.

\begin{acknowledgement}

This material is based upon work supported by the National Science Foundation under Grant No. ECCS-2339271. The authors acknowledge the use of facilities and instrumentation at the UC Irvine Materials Research Institute (IMRI), which is supported in part by the National Science Foundation through the UC Irvine Materials Research Science and Engineering Center (DMR-2011967).

\end{acknowledgement}

\begin{suppinfo}

Details of sample fabrication, diamond anvil cell (DAC) setup, optical characterization, FDTD simulation under pressure, Fano resonance and full width at half maximum (FWHM) analysis, pressure coefficient of refractive index comparison, thickness variation calculation under pressure, estimation of applied pressure in DAC without ruby, and polarization state analysis are provided in the Supporting Information.

\end{suppinfo}


\newpage

\section*{Supporting Information}
\subsection{Section S1. Sample fabrication}
The low-stress free-standing amorphous silicon nitride (SiN) membranes were 5 mm $\times$ 5 mm with 100 nm thickness, purchased from Norcada. Each membrane was supported by a 10 mm $\times$ 10 mm silicon frame. For the Fabry-P\'erot resonators, membranes were coated with 0.4 nm of chromium (Cr), followed by 30 nm of gold (Au) on both sides (as shown in Figure \ref{SI Figure S1}(a)) with a Denton DV-502M Sputter Coating System at the Integrated Nanosystem Research Facility, University of California, Irvine (UCI INRF). 
Figure \ref{SI Figure S1}(b) demonstrates a gold-coated SiN membrane transferred to the diamond surface of a diamond anvil cell system.
The MSs were prepared with a single 30~nm gold coating on top of SiN with Angstrom EvoVac Deposition System at UCI INRF (as shown in Figure \ref{SI Figure S1}(c)). The grating structure is etched solely into the gold layer by Tescan GAIA3 SEM-FIB at UC Irvine Materials Research Institute (UC IMRI) milling with a gallium (Ga) ion source at 30 keV and 30 pA, with a corresponding line dose of 0.01 nC. The designed width of the grating gap is 135 nm, and a period of 565 nm. The total area of the MS is 80 $\mu$m $\times$ 80 $\mu$m.
\setcounter{figure}{0}
\renewcommand{\thefigure}{S\arabic{figure}}
\begin{figure}
  \centering
  \includegraphics[width=1\linewidth]{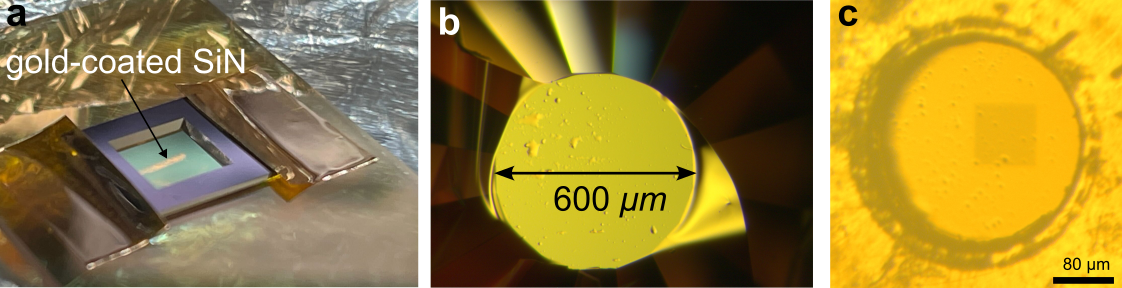}
  \caption{Amorphous silicon nitride (SiN) sample. (a) A gold-coated, free-standing SiN membrane sample enclosed within a 10 mm × 10 mm silicon frame. (b) SiN sample transferred onto a $600\ \mu\text{m}$ diameter anvil culet. (c) The optical images of MS in DAC before the experiment. The gray square presents the metasurface located near to the center of the gasket in the DAC.}
  \label{SI Figure S1}
\end{figure}

\newpage
\subsection{Section S2. Diamond anvil cell (DAC) setup}
High-pressure experiments were conducted in a lever diamond anvil cell (DAC) with a pair of anvils with a culet diameter $d = 600~\mu\text{m}$ (Figure \ref{SI Figure S1}(b)), purchased from Almax-easyLab.
The 250 \(\mu\)m thick stainless steel 304L gaskets were preindented to 90 \(\mu\)m with a 270 \(\mu\)m diameter hole by Almax-easyLab and placed between two diamonds to form a hydrostatic chamber.
The DAC was loaded with methanol:ethanol:water = 16:3:1 mixture solution as the pressure medium to preserve hydrostatic pressure. 
All Fabry-P\'erot resonator experiments were conducted at room temperature up to 5~GPa, at which point the SiN thin films broke. The MS experiments were conducted up to 1.85~GPa.
The exerted pressure was determined by the shift in $R_1$ luminescence of ruby microspheres with diameter of 5 - 10 $\mu$m, excited with a 532 nm laser and calculated by the ruby pressure scale proposed by Mao \textit{et al.} \cite{Mao_Ruby}:
\begin{equation}
    P = \frac{A}{B}\left[\left(\frac{\lambda}{\lambda_0}\right)^B-1\right],
\end{equation}
where $\lambda_0$ is $R_1$ wavelength at ambient pressure (1 atm = 0.1 MPa), and A = 1920~GPa and B = 9.61 are calibration constants \cite{Dewaele_Ruby}. An online tool for ruby fluorescence pressure calculations developed by Kantor was also utilized.\cite{Pressure_Website}

To avoid ruby blocking MS or introducing light scattering, the MS measurements were conducted without ruby. In this case, the applied pressure is estimated from a pressure–rotation relationship established by placing a dial around the rectangular nut of DAC during separate measurements with ruby. 
Details on establishing pressure-rotation relationship are provided in Section S8.

\begin{figure}
  \centering
  \includegraphics[width=0.6\linewidth]{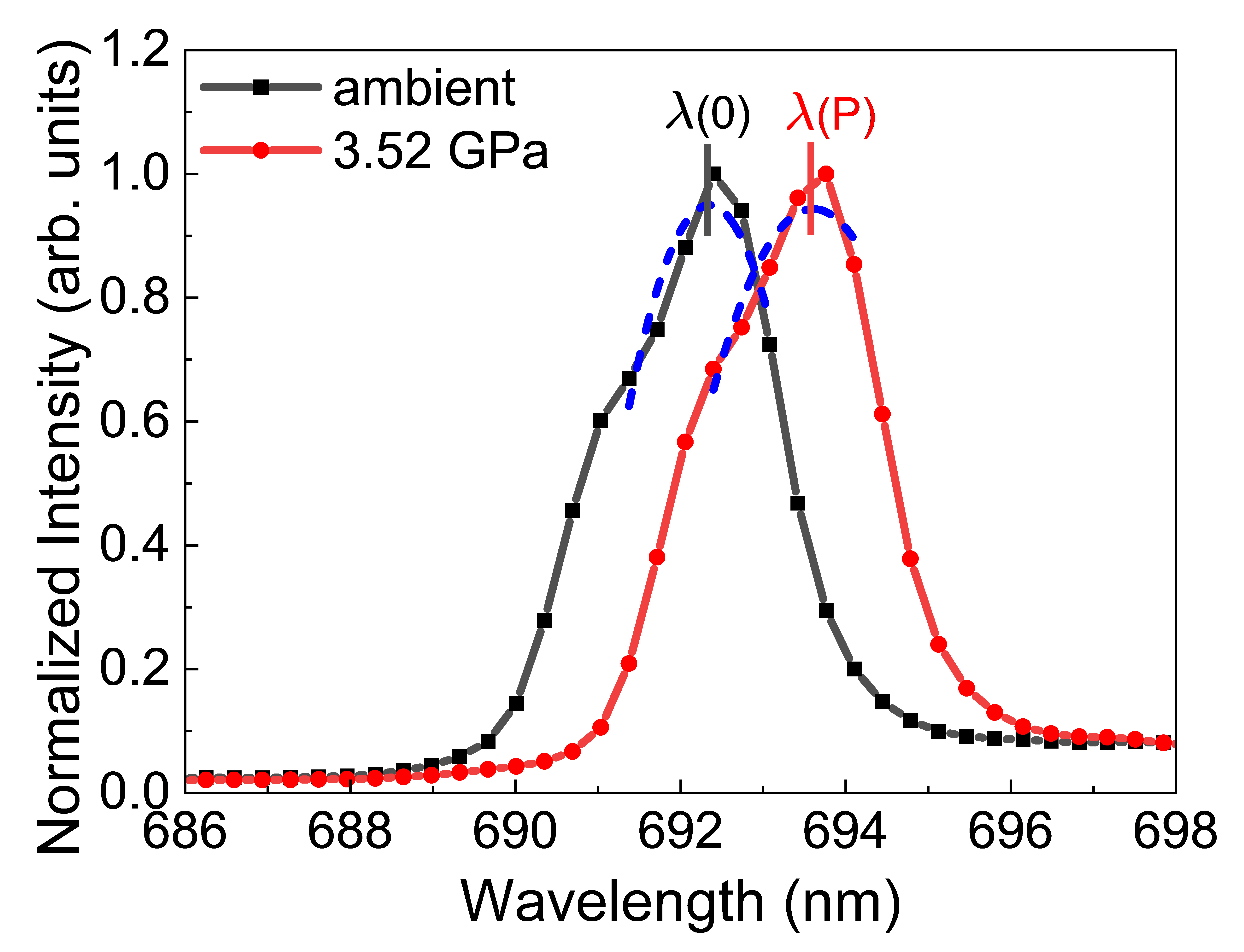}
  \caption{The black and red lines represent the R1 luminescence of the ruby sphere excited by the 532~nm laser at ambient pressure and 3.52~GPa, respectively, showing a shift in ruby luminescence under pressure. Due to the limited resolution of the spectrometer points (black and red points), a second-order polynomial function is implemented to fit the experimental results (the blue dashed lines) and determine the peak positions of the ruby luminescence.}
  \label{SI Figure S2}
\end{figure}
\clearpage

\newpage
\subsection{Section S3. Optical characterization}
\begin{figure}[h]
  \centering
  \includegraphics[width=1\linewidth]{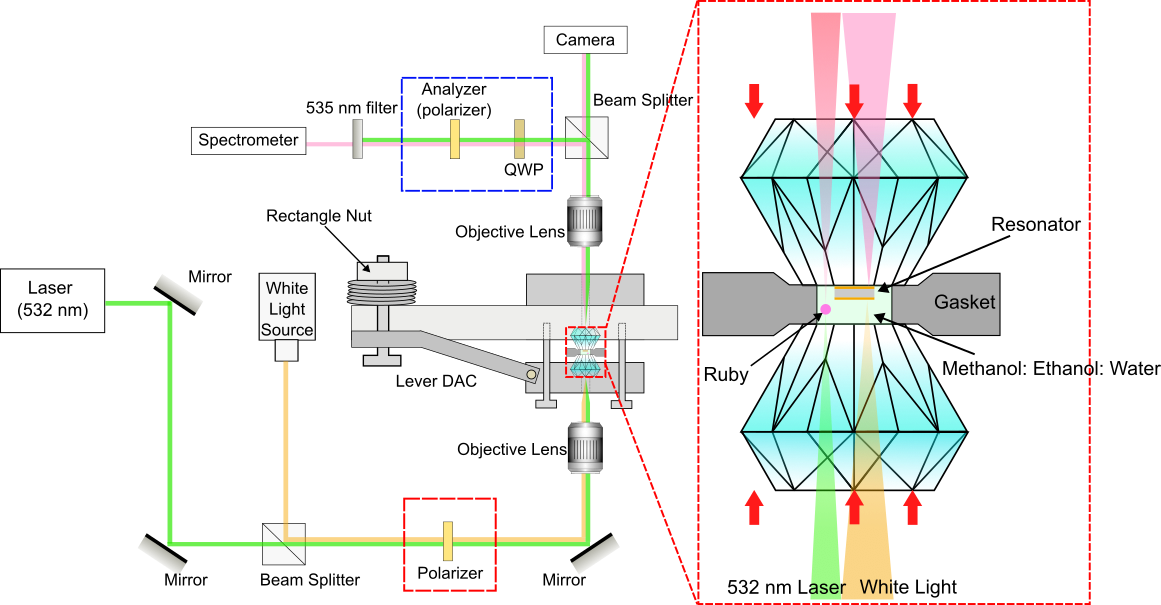}
  \caption{Schematic of extreme pressure diamond anvil cell (DAC) setup.}
  \label{SI Figure S3}
\end{figure}

Figure \ref{SI Figure S3} demonstrates the optical setup with the DAC. All photoluminescence (PL) and optical transmission spectra were collected through a home-built fiber-optic-based microscope equipped with two 10× objective 
with NA = 0.28 and a Ø50 $\mu\text{m}$ multimode fiber with NA = 0.22. The sample and ruby within the DAC were illuminated through an objective by a tungsten light source in the wavelength range of 360–1,000 nm and a 532 nm continuous-wave (CW) solid-state laser, respectively. 
The transmitted light and PL signal were collected from the backside of the DAC using another objective and analyzed by a FLAME-S Ocean Optics miniature spectrometer and a CMOS camera. A 535 nm long-pass filter was employed to suppress the residual 532 nm laser excitation light, thereby enhancing ruby signal detection. 
To ensure all signals originated from the targeted region, we used a Ø50 $\mu\text{m}$ multimode fiber with NA = 0.22 for the Fabry-P\'erot resonator and a Ø25 $\mu\text{m}$ multimode fiber with NA = 0.10 for the MS experiment to connect the light source.
The polarizer, quarter-wave plate (QWP), and analyzer were used in the metasurface experiment to reconstruct and analyze the Stokes parameters of the transmitted light.
By tightening the rectangular nut on the lever DAC, the diamonds are pressed closer to each other and apply pressure to the sample through the pressure medium.

The Stokes parameters are measured as follows: \cite{Shurcliff, AhmedH.Dorrah, MaxBorn, M.Bosch}
\begin{equation}
    S(\lambda, P, \theta', \theta'') =
    \begin{bmatrix}
    S_0  \\
    S_1  \\
    S_2  \\
    S_3  \\
    \end{bmatrix}
    = 
    \begin{bmatrix}
    I(\lambda, P, 0^\circ, \times) + I(\lambda, P, 90^\circ, \times) \\
    I(\lambda, P, 0^\circ, \times) - I(\lambda, P, 90^\circ, \times) \\
    I(\lambda, P, 45^\circ, \times) - I(\lambda, P, -45^\circ, \times) \\
    I(\lambda, P, 0^\circ, 45^\circ) - I(\lambda, P, 0^\circ, -45^\circ) \\
    \end{bmatrix}.
    \label{eq:Stokes parameters}
\end{equation}
where $\theta'$ and $\theta''$ are the angles of the analyzer and QWP, respectively, with respect to the $x$-axis of the DAC frame. We note that the QWP is only inserted for the measurement of $S_3$. The absence of QWP is represented by $\times$ in the formula. 
The orientation $\psi$ and ellipticity $\chi$ of the output elliptical polarization state of the light are calculated as follows: \cite{Shurcliff, AhmedH.Dorrah, MaxBorn, M.Bosch}
\begin{equation}
    \psi = \frac{1}{2} \tan^{-1}\left(\frac{S_2}{S_1}\right), \quad (0^\circ \leq \psi < 180 ^\circ),
\end{equation}
\begin{equation}
    \chi = \frac{1}{2} \sin^{-1}\left(\frac{S_3}{S_0}\right), \quad (-45^\circ \leq \chi \leq 45 ^\circ).
\end{equation}

\newpage

\subsection{Section S4. FDTD simulation under pressure}
\begin{figure}
  \centering
  \includegraphics[width=1\linewidth]{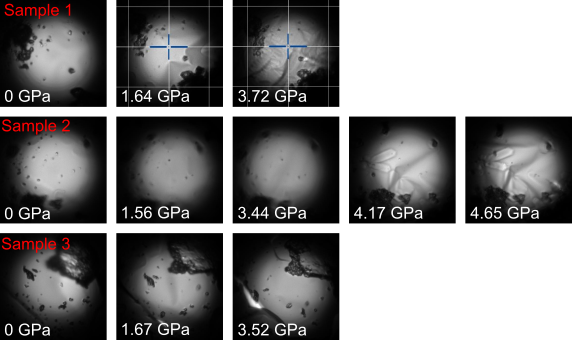}
  \caption{Optical images of the sample under pressure. The images from the top row to the bottom row are sample 1, sample 2, and sample 3, respectively, from ambient pressure to high pressure. We note that all the samples are fully covered in the images. The black particles in the images are ruby powders in the DAC chamber. From ambient pressure to high pressure, the wrinkles on the surface of the sample become larger and begin to compress until they break.}
  \label{SI Figure S4}
\end{figure}

Finite-difference time-domain (FDTD) simulations were performed with the Ansys Lumerical FDTD package to calculate both the Fabry-P\'erot and metasurface resonances and the corresponding electric field distributions. 
The Fabry-P\'erot resonator and the metasurface were simulated with a 20 nm $\times$ 20 nm $\times$ 10 nm mesh and a 5 nm $\times$ 2 nm mesh in 3D and 2D simulations, respectively. 
A frequency-domain field monitor and a power monitor were placed to record the Fabry-P\'erot and metasurface transmittance and electric field distributions. 
A 4,000-nm-wide 2D monitor was placed in the $x-z$ plane to record and calculate the amplitude and phase of the transmitted plane wave after the metasurface structure.
All metasurface simulations are performed using a plane wave light source with polarization angle at $\theta = 45^\circ$ with respect to the grating frame.
Tabulated data were used to model the optical properties of gold \cite{P.B.Johnson} and SiN \cite{LeonidYu.Beliaev}. 
The refractive index of SiN was adjusted by adding 0.02 and 0.16 to the tabulated data to model the non-stoichiometric SiN used in the Fabry-P\'erot and metasurface experiments, respectively. These offsets account for film non-uniformaty arising from wrinkles, and differences between single and double gold layers.
To simulate the pressure-dependent properties of SiN and pressure medium, we adjusted their refractive indices by the relationship obtained from our experiment in Eq. (1) and previous studies in Eq. (2) in the main text.
The wrinkles generated during the sample transfer (as shown in Figure \ref{SI Figure S4}) were simulated by adding a rough surface structure from the Ansys Lumerical FDTD object library on the gold coating in the Fabry-P\'erot resonator with a 5 nm root mean square (RMS) amplitude.  

Considering that the wrinkles on the MS can cause distortions of the grating gap, resulting in either wider or narrower periods and changing the resonance and electric field distributions, we averaged the transmittance and electric field of different grating periods, from 550 nm to 580 nm with 2.5 nm steps, to simulate the influence of wrinkles on the MS.

\newpage
\subsection{Section S5. Fano resonance and FWHM}
\begin{figure}
  \centering
  \includegraphics[width=1\linewidth]{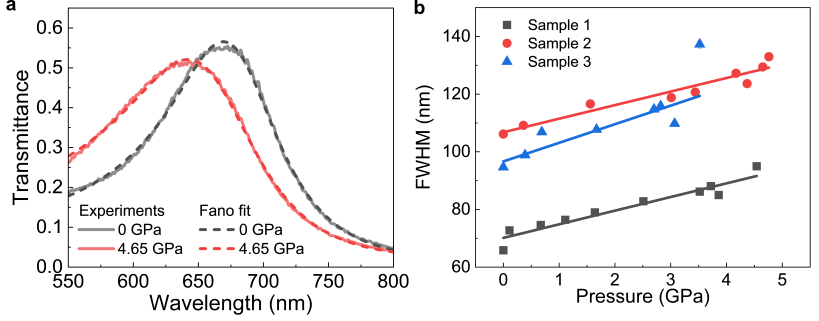}
  \caption{FWHM of transmittance under pressure. (a) The transmittance of the resonance under pressure fitted by Fano functions. The solid lines are the experimental transmittance of the Fabry-P\'erot resonator, and the dashed lines are the Fano functions fitted to the experimental results. (b) FWHM of the transmittance of three samples with respect to pressure. The square, triangular, and circular symbols represent different samples, respectively, and the solid lines are the linear fitting results corresponding to the samples.}
  \label{SI Figure S5}
\end{figure}

To model the non-typical Fabry-P\'erot resonance, we implement Fano function to fit the experimental transmittance of resonator. The Fano function is given by:
\begin{equation}
        \label{Fano equation}
        y = y_0 + \frac{H \left( 1 + \frac{x - x_c}{q \omega} \right)^2}{1 + \left( \frac{x - x_c}{q \omega} \right)^2},
\end{equation}
where \(y_0\) is the baseline offset, \(x_c\) is the resonance center, H is the peak amplitude, \(\omega\) is the resonance width, and q represents the asymmetry parameter.
The Fano fitting results are depicted in Figure \ref{SI Figure S5}a. The excellent agreement between the experimental transmittance and the Fano function indicates that the coupling of the Fabry-P\'erot resonance and the background light gives rise to the Fano-like profile observed in the transmittance results \cite{MikhailF.Limonov}.

With Fano parameter, we analyze the full width at half maximum (FWHM) of the resonator with increasing pressures (as shown in Figure \ref{SI Figure S5}b). The FWHM of three resonators increase on average by nearly 25 nm under pressure, and the derivatives of the FWHM with respect to pressure across three resonators are approximately 4.85 $\pm$ 0.3 nm/GPa, demonstrating the resonance consistently broaden when the pressure increases. These broadened resonances might result from the larger wrinkles on the samples, as shown in Figure \ref{SI Figure S4}.

\newpage
\subsection{Section S6. Pressure coefficient of refractive index comparison}
\begin{table} 
  \small 
  \centering
  \renewcommand{\arraystretch}{1.5}
  \begin{tabular}{ c c c c c}
    \hline
    \hline
        Material & $B_0$ (GPa) & $B'_0$ &  $\frac{dn}{dP}$ (\(10^{-2}\) GPa$^{-1}$) & Ref. \\
    \hline
        SiN & 290\cite{Youngs_Modulus} & 4\cite{Derivative_of_Modulus}  & \(-(1.5\pm0.16)\) & This work, transmission, DAC\\
        
        Ge & 74.4 & 4.76 & \(-(4.5)\) & transmission, DAC\cite{Ge}\\
 
        c-BN & 382 & 4.1  & \(-(0.06\pm0.6)\) & transmission, DAC\cite{diamond-SiC-cBN}\\
        
        3C-SiC & 227 & 4.1  & \(-(0.23\pm0.5)\) & transmission, DAC\cite{diamond-SiC-cBN}\\
        
        Diamond & 442 & 4  & \(-(0.08\pm0.2)\) & transmission, DAC\cite{diamond-SiC-cBN}\\

        GaN & 200 & 4.3  & \(-(0.7)\) & transmission, DAC\cite{ZnO}\\

        AIN & 208 & 3.6  & \(-(0.17)\) & transmission, DAC\cite{ZnO}\\

        GaAs & 74.7 & 4.46 & \(-(1.3)\) & transmission, DAC\cite{Ge}\\
    \hline
    \hline
  \end{tabular}
  \caption[Experimentally derived parameters of the pressure-dependent refractive index.] {Experimentally derived parameters of the pressure-dependent refractive index. The bulk modulus (\(B_0\)) of SiN presented in this work was derived from the Young’s modulus reported by Khan \textit{et al.} \cite{Youngs_Modulus}. The pressure derivative of the bulk modulus (\(B_0'\)) was calculated as the average of the values reported for different phases of silicon nitride in the study by Xu \textit{et al.} \cite{Derivative_of_Modulus}.}
  \label{tab:Pressure coefficient of refractive index}
\end{table}
\newpage

\subsection{Section S7. Calculation of thickness variation under pressure}

\begin{figure}
  \centering
  \includegraphics[width=.95\linewidth]{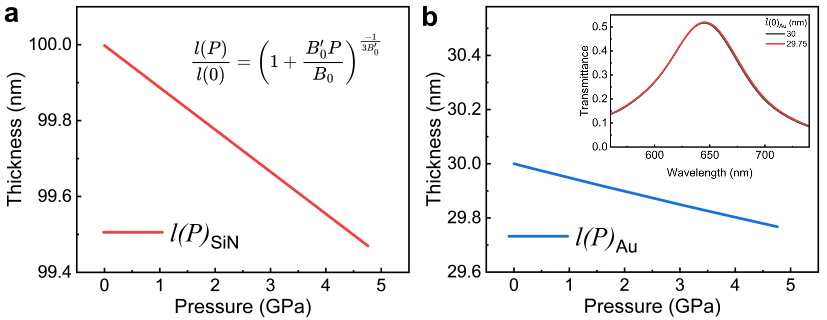}
  \caption{Thickness variation of SiN and gold under pressure. The thickness variation is calculated by Murnaghan’s first order equation \cite{MURNAGHAN,MgO}. \textbf{a,} The thickness variation of SiN under pressure. By calculation, the thickness of SiN decreases from 100 nm to 99.47 nm (0.55\% decrease) under 4.76 GPa, demonstrating that the variation of SiN's thickness under pressure is limited. \textbf{b,} The thickness variation of gold under pressure. The inset shows the FDTD simulation result of the resonator with varying gold thickness on both sides. The absence of a resonance shift confirms that the thickness variation of gold is not the primary factor in resonance shift.}
  \label{SI Figure S6}
\end{figure}

Figure \ref{SI Figure S6}a and \ref{SI Figure S6}b show the thickness variation of SiN and gold under pressure, respectively, calculated using Murnaghan’s first order equation \cite{MURNAGHAN,MgO}. 
The bulk modulus of SiN $B_0 = 290$~GPa was derived from the Young’s modulus reported by Khan \textit{et al.} \cite{Youngs_Modulus}, and the corresponding pressure derivative of the bulk modulus $B_0'=6$ was calculated as the average of the values reported for different phases of silicon nitride in the study by Xu \textit{et al.}
The parameters used are $B_0 = 190$~GPa and $B'_0 = 4$ for gold \cite{C.Martin-Sanchez, C.Martin-Sanchez2}.
The variation of SiN’s thickness is only 0.55\% under pressure; in contrast, the shift of the resonance exceeds 4.5\%, demonstrating that the shift is primarily dominated by the change in the refractive index rather than thickness variation.
Although the thickness variation of gold is 0.83\%, there is almost no corresponding resonance shift (as shown in the FDTD simulation result in the inset of Figure\ref{SI Figure S6}b), further indicating that the variation of metal thickness is not the main factor in the resonance shift. This finding indicates that the shift of the resonance mainly comes from variation in the refractive index of cavity medium.

\newpage

\newpage
\subsection{Section S8. Estimation of applied pressure in DAC without ruby}
\begin{figure}
  \centering
  \includegraphics[width=1\linewidth]{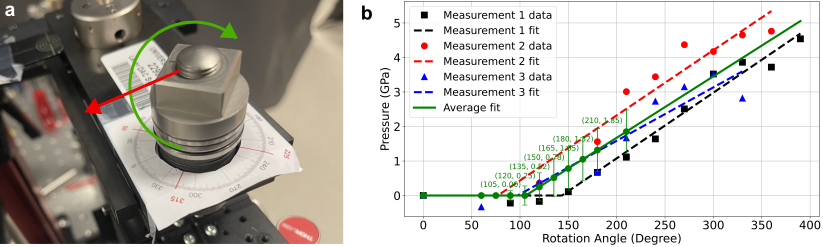}
  \caption{Method for estimating applied pressure in the DAC without ruby. (a) The dial is placed around the rectangular nut of the DAC to establish the pressure–rotation relationship. The corner of the nut is aligned to 0$^\circ$, as indicated by the red arrow, defining the ambient pressure. The green arrow shows the rotation direction for increasing pressure in the DAC. (b) The relationship between the rotation angle of the nut and the applied pressure in the DAC. The square, circular, and triangular symbols represent three different measurements of pressure with ruby in the DAC. By taking an average of these results and fitting with a ReLU function, the average represents the general relationship between the rotation of the nut and the applied pressure in the DAC. In general, the pressure starts to increase when the nut is rotated beyond 105$^\circ$.}
  \label{SI Figure S7}
\end{figure}

To estimate the applied pressure in the DAC without ruby, we establish a pressure–rotation relationship by placing a dial around the rectangular nut of DAC during a measurement with ruby, as shown in Figure \ref{SI Figure S7}a.

For calibration, we measure the pressure with ruby in DAC. As indicated by the red arrow in Figure \ref{SI Figure S7}a, we align the corner of the rectangular nut to 0$^\circ$ on the dial and define it as the ambient pressure. By rotating the rectangular nut to increase the pressure, the relationship between rotation angle and applied pressure is obtained, as shown in Figure \ref{SI Figure S7}b. An average of all the pressure measurements is used to generalize the pressure–rotation relationship and is fitted with a ReLU function, as indicated by the green line. The result shows that pressure starts to increase when the rectangular nut rotates beyond 105$^\circ$. With this relationship, we are able to estimate the applied pressure in the DAC when ruby is not used.

\newpage
\subsection{Section S9. Metasurface before and after high-pressure experiment} 
Figure S8 shows the optical image of the metasurface after the extreme pressure experiment. The remaining grating structure of the metasurface indicates its pressure-tunability is reversible up to 1.85~GPa. This finding shows the potential for barophotonic metasurfaces as a platform for optical applications with various designs.

\begin{figure}
  \centering
  \includegraphics[width=0.5\linewidth]{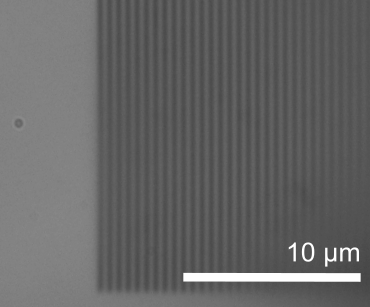}
  \caption{Optical image of the metasurface after extreme pressure experiment. The preserved grating structure demonstrates that the device maintains reversibility up to 1.85~GPa.}
  \label{SI Figure S8}
\end{figure}

\newpage
\subsection{Section S10. FDTD simulation for wrinkles on metasurface}
\begin{figure}
  \centering
  \includegraphics[width=1\linewidth]{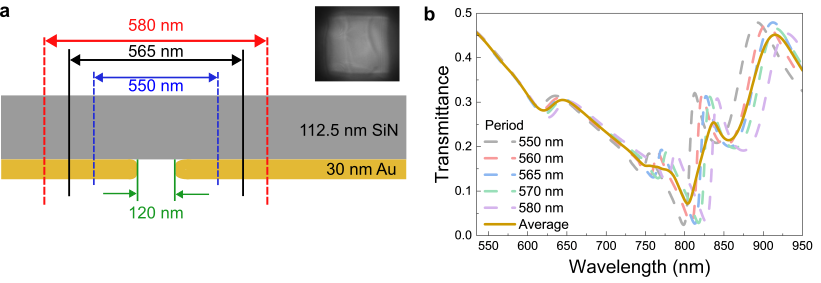}
  \caption{FDTD for wrinkles on metasurface. (a) The illustration of the FDTD region varies from 15 nm wider (550 nm, the blue dashed lines region) to 15 nm narrower (580 nm, the red dashed lines region) than the actual grating period (565 nm, the black solid lines region), simulating distortions of the grating gap caused by the wrinkles on the metasurface. The green solid lines show the gap of the grating width, which is 120 nm wide. The thickness of the SiN and the gold coating are 112.5 nm and 30 nm, respectively. These simulation dimensions account for the fabrication tolerance. The inset shows the optical image of the metasurface in the DAC, demonstrating the wrinkles on the metasurface. (b) The FDTD transmittance of the metasurface with different grating gap periods. The black, red, blue, green, and purple dashed lines represent the transmittance for the grating periods of 550, 560, 565, 570, and 580 nm, respectively. The yellow solid line is the average transmittance over the full range from 550 to 580 nm in 2.5 nm steps.}
  \label{SI Figure S9}
\end{figure}

Some wrinkles are generated during the sample transfer onto diamond, as shown in the inset of Figure \ref{SI Figure S9}a. These wrinkles on the metasurface can cause slight distortions of the periods of the grating gap, resulting in either wider or narrower periods and changing the resonance and electric field distributions. To account for the influence of wrinkles in simulation, 
we averaged the transmittance and electric field of different grating periods, from 550 nm to 580 nm with 2.5 nm steps, to simulate the influence of wrinkles on the MS, as depicted in Figure \ref{SI Figure S9}a. 
Figure \ref{SI Figure S9}b shows the transmittance of period of the gaps with 550, 560, 565 (the same dimension used in experiment), 570, and 580 nm, as well as the average transmittance over the full range from 550 to 580 nm in 2.5 nm steps. By averaging the transmittance of the periods that are 15 nm wider and narrower than the actual grating period used in experiment, the influence of wrinkles on the metasurface is relatively simulated in FDTD.

\newpage
\subsection{Section S11. Output polarization state calculation}
\begin{figure}
\centering
\includegraphics[width=1\linewidth]{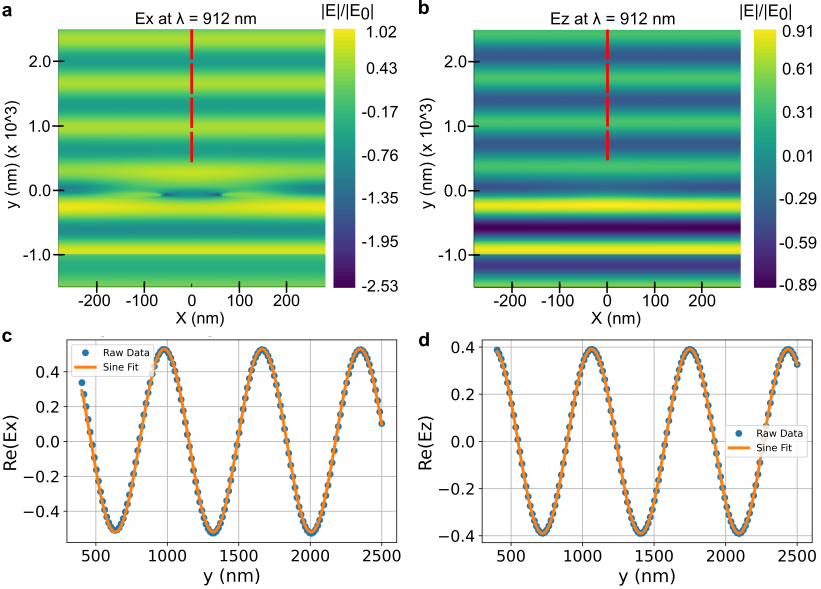}
\caption{Method for analyzing the output polarization states of light in FDTD. (a)-(b) The $x$-component and $z$-component of the electric field, $E_x$ and $E_z$, respectively, at $\lambda = 912$ nm. The red dashed lines indicate the region at $x = 0$ nm and $y = 400$ to 2500 nm, where $E_x$ and $E_z$ are stable after passing through the sample. (c)-(d) The $E_x$ and $E_z$ extracted along the red dashed lines in (a) and (b) are fitted with sine functions.}
\label{SI Figure S10}
\end{figure}

All the simulations are performed using a linearly polarized incident light at $45^\circ$ with respect to the grating metasurface frame, over a wavelength range from $\lambda = 500$ to 950 nm, in a 2-dimensional FDTD simulation. Figure \ref{SI Figure S10}a and \ref{SI Figure S10}b show the $x$-component and $z$-component of the electric field, $E_x$ and $E_z$, respectively, at $\lambda = 912$ nm. 

To analyze the output polarization states, $E_x$ and $E_z$ are extracted at $x$ = 0 nm, from $y=$ 400 to 2500 nm (as indicated by the red dashed lines in Figure \ref{SI Figure S10}a and \ref{SI Figure S10}b), in the region where the electric field becomes stable after passing through the sample. The results are shown in Figure \ref{SI Figure S10}c and \ref{SI Figure S10}d for the $E_x$ and $E_z$ at $\lambda = 912$ nm, respectively, and are fitted using sine functions. The amplitudes and phases of $E_x$ and $E_z$ are determined from the fitting parameters. 

The orientation angle $\psi$ and the ellipticity angle $\chi$ of the output polarization states were calculated by the equations below \cite{MaxBorn, Shurcliff, AhmedH.Dorrah, M.Bosch}:
\begin{equation}
    \tan{2\psi} = \frac{2E_x E_z}{E_x^2 - E_z^2} \cos{\phi}, \quad (0 \leq \psi < \pi);
\end{equation}
\begin{equation}
    \sin{2\chi} = \frac{2E_x E_z}{E_x^2 + E_z^2} \sin{\phi}, \quad (-\frac{\pi}{4}\leq \chi \leq \frac{\pi}{4}),
\end{equation}
where $E_x$ and $E_z$ are the amplitudes of $x$-component and $z$-component of the electric field, respectively, and $\phi = \phi_x-\phi_z$, which is the phase difference between $E_x$ and $E_z$.

\newpage

\providecommand{\latin}[1]{#1}
\makeatletter
\providecommand{\doi}
  {\begingroup\let\do\@makeother\dospecials
  \catcode`\{=1 \catcode`\}=2 \doi@aux}
\providecommand{\doi@aux}[1]{\endgroup\texttt{#1}}
\makeatother
\providecommand*\mcitethebibliography{\thebibliography}
\csname @ifundefined\endcsname{endmcitethebibliography}
  {\let\endmcitethebibliography\endthebibliography}{}

\end{document}